\documentclass[letterpaper, 10 pt, conference]{ieeeconf}  

\IEEEoverridecommandlockouts                              
\usepackage{graphicx}      
\usepackage{subcaption}
\usepackage{siunitx}
\usepackage{amsmath}
\usepackage{amssymb}
\usepackage{adjustbox}
\usepackage{parskip} 
\usepackage{tikz}
\usepackage{pgfplots}
\pgfplotsset{compat=1.17}
\usepackage[version=4]{mhchem}

\definecolor{Tblue}{HTML}{1f77b4}
\definecolor{Torange}{HTML}{ff7f0e}
\definecolor{Tgreen}{HTML}{2ca02c}
\definecolor{Tred}{HTML}{d62728}
\definecolor{Tpurple}{HTML}{9467bd}
\definecolor{Tbrown}{HTML}{8c564b}

\makeatletter
\newenvironment{customlegend}[1][]{%
    \begingroup
    \pgfplots@init@cleared@structures
    \pgfplotsset{#1}%
}{
    \pgfplots@createlegend
    \endgroup
}

\def\addlegendimage{\pgfplots@addlegendimage}

\newcommand{\state}[0]{\mathbf{x}}
\newcommand{\xinput}[0]{\mathbf{u}}
\newcommand{\source}[0]{\mathbf{f}}

\newcommand{\net}[0]{\hat{\mathbf{f}}}

\newcommand{\timestep}[0]{{h}}

\DeclareMathOperator {\std}{std}

\usepackage[nolist,nohyperlinks]{acronym}
\acrodef{NN}{neural network}
\acrodef{PGNN}{physics-guided neural network}
\acrodef{PINN}{physics-informed neural network}
\acrodef{PBM}{physics-based models}
\acrodef{DDM}{Data-driven modeling}
\acrodef{HAM}{hybrid analysis and modeling}
\acrodef{ML}{machine learning}
\acrodef{MSE}{mean squared error}
\acrodef{RFMSE}{rolling forecast mean squared error}
\acrodef{MAV}{miniature air vehicle}
\acrodef{PDE}{partial differential equation}
\acrodef{ODE}{ordinary differential equations}
\acrodef{RNN}{recurrent neural network}
\acrodef{ROM}{reduced order model}
\acrodef{DOF}{degrees of freedom}
\acrodef{CG}{centre of gravity}
\acrodef{PCR}{principal component regression}
\acrodef{PCA}{principal component analysis}
\acrodef{NED}{North-East-Down}
\acrodef{PID}{Proportional-Integral-Derivative}
\acrodef{NN}{neural network}
\acrodef{PWA}{piecewise affine}
\acrodef{MILP}{mixed integer linear programming problem}
\acrodef{SMT}{satisfiability modulo theory}
\acrodef{IVP}{initial value problem}
\acrodef{MPC}{model predictive control}
\acrodef{PWL}{piecewise linear}
\acrodef{CFD}{computational fluid dynamics}
\acrodef{EKF}{Extended Kalman Filter}
\acrodef{RK4}{Runge-Kutta 4}
\acrodef{CoSTA}{Corrective source term approach}
\acrodef{APRBS}{Amplitude-modulated Pseudo-Random Binary Signal}
\acrodef{SciML}{Scientific Machine Learning}
\acrodef{ACD}{Anode-Cathode Distance}
\acrodef{UAV}{unmanned aerial vehicle}
\acrodef{PE}{persistency of excitation}
\acrodef{RL}{Reinforcement Learning}

\title{\LARGE \bf
Sparse neural networks with skip-connections for identification of aluminum electrolysis cell
}

\author{Erlend Torje Berg Lundby, Haakon Robinson, Adil Rasheed, Ivar Johan Halvorsen, Jan Tommy Gravdahl
\thanks{This work was supported by the project TAPI: Towards Autonomy in Process Industries (grant no. 294544), and EXAIGON: Explainable AI systems for gradual industry adoption (grant no. 304843)}
\thanks{E.T.B. Lundby, H. Robinson, A. Rasheed, and J. T. Gravdahl are with the Department of Engineering Cybernetics,
        Norwegian University of Science and Technology, 7034 Trondheim, Norway
        {\tt\small <erlend.t.b.lundby, haakon.robinson, adil.rasheed, jan.tommy.gravdahl>@ntnu.no}}%
\thanks{I. J. Halvorsen is with SINTEF Digital, Trondheim, No-7465, Norway
        {\tt\small ivar.j.halvorsen@sintef.no}}%
}

\begin{document}

\maketitle
\thispagestyle{empty}
\pagestyle{empty}

\begin{abstract}
Neural networks are rapidly gaining interest in nonlinear system identification due to the model's ability to capture complex input-output relations directly from data.
However, despite the flexibility of the approach, there are still concerns about the safety of these models in this context, as well as the need for large amounts of potentially expensive data. Aluminum electrolysis is a highly nonlinear production process, and most of the data must be sampled manually, making the sampling process expensive and infrequent. In the case of infrequent measurements of state variables, the accuracy and open-loop stability of the long-term predictions become highly important. Standard neural networks struggle to provide stable long-term predictions with limited training data. In this work, we investigate the effect of combining concatenated skip-connections and the sparsity-promoting $\ell_1$ regularization on the open-loop stability and accuracy of forecasts with short, medium, and long prediction horizons. The case study is conducted on a high-dimensional and nonlinear simulator representing an aluminum electrolysis cell's mass and energy balance. The proposed model structure contains concatenated skip connections from the input layer and all intermittent layers to the output layer, referred to as InputSkip. $\ell_1$ regularized InputSkip is called sparse InputSkip. The results show that sparse InputSkip outperforms dense and sparse standard feedforward neural networks and dense InputSkip regarding open-loop stability and long-term predictive accuracy. The results are significant when models are trained on datasets of all sizes (small, medium, and large training sets) and for all prediction horizons (short, medium, and long prediction horizons.)
\end{abstract}

\section{Introduction}

There is increasing interest in using machine learning-based methods to develop predictive models directly from data.
Compared to standard system identification methods, the advantage of data-driven modeling is that it does not require any assumptions about the system. However, all phenomena well represented by the data can often be captured accurately. 
One example of such a method that has seen widespread popularity in recent years is the \ac{NN}, which is known to be a universal function approximator. 
These are often used in \ac{RL} to represent a value function or a model for some dynamical system.
However, this approach requires many data points to train effective models, which can be expensive in many domains.
One hypothesis is that \ac{NN}s are typically overparameterized and require many steps to adjust all parameters.
However, overtraining on the same limited dataset will cause the model to overfit the training data and perform poorly on unseen data.
While overparameterization has been found to aid convergence during training \cite{allenzhu_convergence_2019}, it also introduces redundant information into the weights.

Recent research has found that sparser networks may be the key to training models that generalize across many situations.
In particular, it has been shown empirically that for any dense architecture, there is a high probability that there is a sparse subnetwork that will train faster and generalize better than the full model \cite{frankle2018the}. 
This phenomenon is known as the Lottery Ticket Hypothesis. Many sparsification methods can be seen as attempts to extract such a ``winning lottery ticket" from an initially dense network.
In system identification, previous work shows that sparsity-promoting $\ell_1$ regularization can benefit model generalization, interpretability, and stability \cite{lundby2022sdn}.
Group sparsity methods have also been applied to Bayesian recurrent neural networks, with favorable results \cite{Zhou2022sbd}. 
There have been numerous advances in this field, and we refer to \cite{hoefler2021sparsity} for a recent and comprehensive review.
This work uses the well-known $\ell_1$ regularization to induce sparsity in neural networks.

Another challenge related to using \ac{NN}s is the choice of architecture and hyperparameters.
Typical networks have multiple layers which are densely connected, although this can vary between domains.
Choosing an appropriate architecture is an art involving trial and error to improve performance and avoid overfitting.
It is commonly understood that the early layers of a neural network significantly impact the overall performance of the network. However, deep networks often suffer from the vanishing or exploding gradient problem, which prevents effective training of these early parameters \cite{goodfellow2016deep}.
Skip-connections were originally proposed to circumvent this by introducing a shorter path between the early layers and the output \cite{He2015}.
They were found to enable the training of significantly deeper networks but may also improve training convergence \cite{li2017vll}. 


In the dynamical systems and control field, models are often designed with a purpose in mind, such as designing a control system or state observer.
Crucially, we are interested in the behavior and performance of the controlled system regarding objectives such as energy efficiency or yield, implying that the model does not need to be perfectly accurate for the entire state space so long as the resulting closed-loop performance is sufficient (known as \textit{identification for control} (I4C)).
If high-frequency measurements from the system are available, only the short-term behavior of the model is important since any drift out of the operational space is quickly corrected.
However, if measurements are rarely available, such as in the aluminum electrolysis process that we consider, the long-term model behavior and open-loop stability become much more critical. 
Stable long-term predictions can be important for decision-making, meaning a model with good long-term stability and accuracy is inherently meaningful.

In this work, we investigate the effects of adding skip connections and $\ell_1$ regularization on the accuracy and stability of these models for short, medium, and long horizons. The following questions are addressed:
\begin{itemize}
    \item How do skip connections affect the stability and generalization error of neural networks trained on high-dimensional nonlinear dynamical systems? 
    \item How does sparsity affect stability and generalization error for neural networks with skip connections when modeling nonlinear dynamics?
    \item How does the amount of training data affect neural networks with skip connections compared to neural networks without skip connections?
\end{itemize}
We make the following contributions:
\begin{itemize}
    \item We perform a black box system identification of an aluminum electrolysis cell using different \ac{NN} architectures.
    \item We demonstrate that the accuracy and open-loop stability of the resulting models is greatly improved by using $\ell_1$ weight regularization and incorporating skip connections into the architecture.
    \item This advantage is consistent across datasets of varying sizes.
\end{itemize}

\section{Theory}
\subsection{Physics-based model for aluminum extraction}
\label{subsec:pbm}
\begin{figure}
\begin{center}
\includegraphics[width=\linewidth]{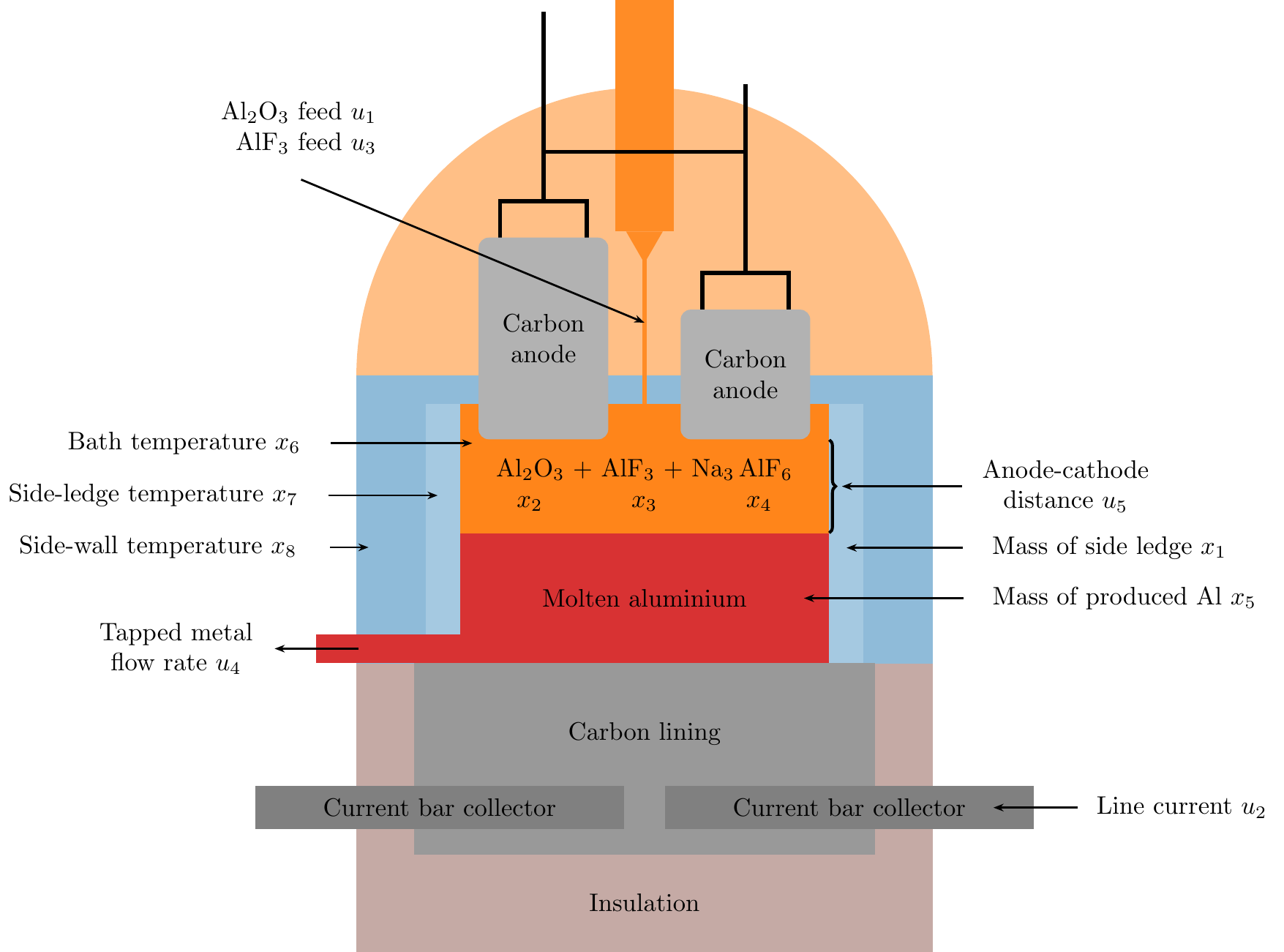}    
    \caption{Schematic of the setup}
    \label{fig:schematicofthesetup}
\end{center}
\end{figure}
\ac{NN}s are first trained on synthetic data generated from a known \ac{PBM}.
The model used in this work describes the internal dynamics of an aluminum electrolysis cell based on the Hall-H{\'e}roult process. 
Fig.~\ref{fig:schematicofthesetup} shows a diagram of the electrolysis cell. 
Traditional \ac{PBM}s of such systems are generally constructed by studying the mass/energy balance of the chemical reactions.
The system is described by a set of \ac{ODE}:
\begin{equation}
    \Dot{\state} = \source(\state, \xinput),
    \label{eq:nonlin_state_space}
\end{equation}
where $\state \in \mathbb{R}^8$ and $\xinput \in \mathbb{R}^5$ represent the time-varying states and inputs of the system respectively.
The full set of equations are:
\begin{subequations}\label{eq:alu_equations}
\begin{align}
        \Dot{x}_1 &=\,\frac{k_1(g_1 - x_7)}{x_1 k_0} - k_2 (x_6 - g_1) \\
        \Dot{x}_2 &=\, u_1 - k_3 u_2\\
        \Dot{x}_3 &=\, u_3 - k_4 u_1\\
        \Dot{x}_4 &=\, -\frac{k_1 (g_1 - x_7)}{x_1 k_0} + k_2 (x_6 - g_1) + k_5 u_1\\
        \Dot{x}_5 &=\, k_6 u_2 - u_4 \\
        \Dot{x}_6 &=\, \frac{\alpha}{x_2+x_3+x_4} \Bigg[ u_2 g_5 + \frac{u_2^2 u_5}{2620 g_2} - k_7 (x_6 - g_1)^2 \\
                  & \, \;  + k_8 \frac{(x_6 - g_1)(g_1 - x_7)}{k_0x_1} - k_9 \frac{x_6 - x_7}{k_{10} + k_{11} k_0  x_1} \Bigg] \nonumber\\
        \Dot{x}_7 &=\, \frac{\beta}{x_1} \biggl[ \frac{k_{9} (g_1 - x_7)}{k_{15}k_0 x_1} - k_{12}(x_6 - g_1)(g_1 - x_7)  \\
                  & \, \; +  \frac{k_{13}(g_1 - x_7)^2}{k_0x_1} 
                  \,  - \frac{x_7 - x_8}{k_{14} + k_{15}k_0  x_1} \biggr]\nonumber\\
        \Dot{x}_8 &=\, k_{17} k_9 \left(\frac{x_7 - x_8}{k_{14} + k_{15} k_0 \cdot x_1} - \frac{x_8 - k_{16}}{k_{14} + k_{18}}\right),
\end{align}
\end{subequations}
 
where the intrinsic properties $g_i$ of the bath mixture are given as:

\begin{subequations}\label{eq:alu_nonlin_help_fun}
\begin{align}
    g_1 &=  991.2 + 112 c_{x_3} + 61 c_{x_3}^{1.5} - 3265.5 c_{x_3}^{2.2} \label{eq:liquidus_temp}  \\
        &- \frac{793 c_{x_2}}{- 23 c_{x_2} c_{x_3} - 17 c_{x_3}^{2} + 9.36 c_{x_3} + 1} \nonumber \\
    g_2 &=  \text{exp}\,\left(2.496 - \frac{2068.4}{273+x_6} - 2.07c_{x_2}\right)\\
    g_3 &= 0.531 + 3.06 \cdot 10^{-18} u_{1}^{3} - 2.51 \cdot 10^{-12} u_{1}^{2} \\ &+ 6.96 \cdot 10^{-7} u_{1} 
        - \frac{14.37 (c_{x_2} - c_{x2,crit}) - 0.431}{735.3 (c_{x_2} - c_{x2,crit}) + 1}  \nonumber\\
    g_4 &= \frac{0.5517 + 3.8168 \cdot 10^{-6}u_2}{1 + 8.271 \cdot 10^{-6}u_2}\\
    g_5 &= \frac{3.8168\cdot 10^{-6} g_3 g_4 u_2}{g_2(1 -g_3)}. 
\end{align}       
\end{subequations}
See Table~\ref{table:states_inputs} for a description of these quantities.
The dynamics of the system are relatively slow. The control inputs $u_1, \; u_3$ and $u_4$ are therefore well modeled as impulses representing discrete events involving the addition or removal of substances. This results in step changes in the linear states $x_2, x_3, x_5$, which act as accumulator states for the mass of the corresponding substance (see Table~\ref{table:states_inputs}). The control inputs $u_2$ and $u_5$ are piecewise constant and nonzero. The inputs $\xinput$ are determined by a simple proportional controller $\boldsymbol{\pi}(\state)$. The simulation model is derived in \cite{lundby2022sdn}, and we refer to that article for the values of the simulation parameters and further details.
\begin{table}
    \centering
    \caption{Table of states, inputs, and other quantities used to model the electrolysis cell}
    \label{table:states_inputs}
    \small
    \begin{tabular}{c|l|c}
    \hline
    Variable & Physical meaning & Units  \\ \hline
    $x_1$ & Mass side ledge & $\si{kg}$ \\
    $x_2$ & Mass \ce{Al2O3} & $\si{kg}$  \\
    $x_3$ & Mass \ce{AlF_3} & $\si{kg}$ \\
    $x_4$ & Mass \ce{Na_3 AlF_6} & $\si{kg}$  \\
    $x_5$ & Mass metal & $\si{kg}$ \\
    $x_6$ & Temperature bath & $\si{\degreeCelsius}$ \\
    $x_7$ & Temperature side ledge & $\si{\degreeCelsius}$ \\
    $x_8$ & Temperature side wall & $\si{\degreeCelsius}$\\
    \hline
    $u_1$ &  \ce{Al2O3} feed &$\si{kg/s}$ \\
    $u_2$ & Line current  & $\si{kA}$\\
    $u_3$ & \ce{AlF_3} feed & $\si{kg/s}$\\
    $u_4$ & Aluminum tapping & $\si{kg/s}$\\
    $u_5$ & Anode-cathode distance & $\si{cm}$ \\
    \hline
    $c_{x_2}$ & \ce{Al2O3} mass ratio $x_2/(x_2 + x_3 + x_4)$  & - \\
    $c_{x_3}$ & \ce{AlF_3} \ \ mass ratio $x_3/(x_2 + x_3 + x_4)$ & - \\
    \hline
    $g_1$ & Liquidus temperature    & $\si{\degreeCelsius}$    \\
    $g_2$ & Electrical conductivity & $\si{\siemens\meter}$    \\
    $g_3$ & Bubble coverage         & -    \\
    $g_4$ & Bubble thickness        & $\si{cm}$    \\
    $g_5$ & Bubble voltage          & $\si{\volt}$    \\
    \hline
    \end{tabular}
\end{table}

\subsection{Deep neural network with skip connections}
\label{subsec:skip_con}

\begin{figure}
\begin{center}
\includegraphics[width=\linewidth]{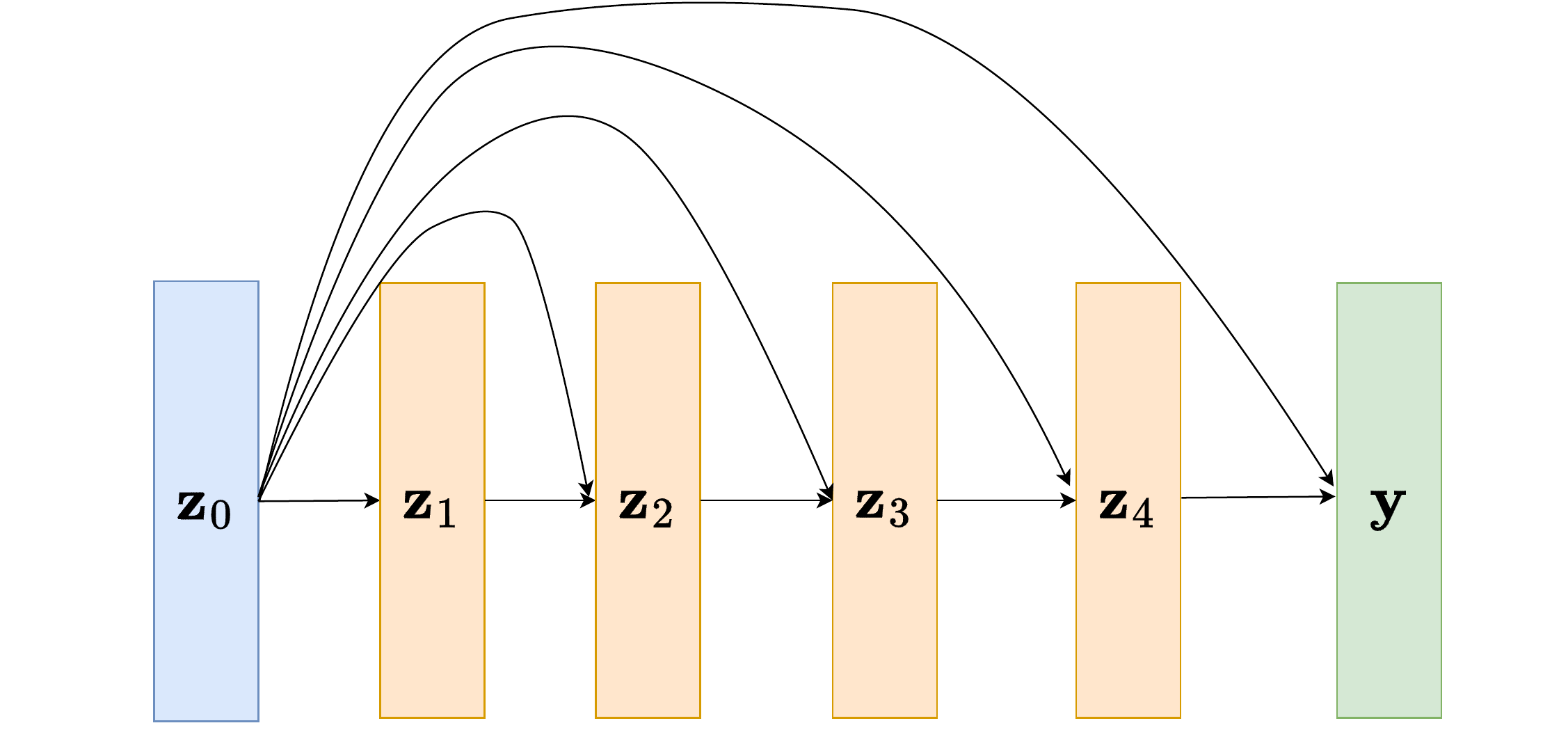}    
    \caption{InputSkip architecture with 4 hidden layers}
    \label{fig:inputSkip}
\end{center}
\end{figure}

A \ac{NN} with $L$ layers can be compactly written as an alternating composition of affine transformations $\mathbf{Wz + b}$ and nonlinear activation functions $\boldsymbol{\sigma} : \mathbb{R}^n \mapsto \mathbb{R}^n$:
%
%
\begin{equation} 
\begin{aligned}
    \mathbf{z}_{i} = \boldsymbol{\sigma}_i\left(\mathbf{W}_i\mathbf{z}_{i-1} + \mathbf{b}_i\right)
\end{aligned}
\end{equation}
where $\mathbf{z}_0$ is the input to the network, the activation function $\boldsymbol{\sigma}_i$, weight matrix $\mathbf{W}_i$, and bias vector $\mathbf{b}_i$ correspond to the $i$th layer of the network. The universal approximation property of \ac{NN}s makes them very attractive as a flexible model class when a lot of data is available. The representation capacity is generally understood to increase with both the depth and the width (the number of neurons in each layer), although early attempts to train very deep networks found them challenging to optimize using backpropagation due to the vanishing gradients problem. 
One of the major developments that enabled researchers to train deep \ac{NN}s with many layers is the \textit{skip connection}. A skip connection is simply an additional inter-layer connection that bypasses some of the layers of the network. This provides alternate pathways through which the loss can be backpropagated to the early layers of the NN, which helps mitigate the issues of vanishing and exploding gradients, which were major hurdles to training deeper models. In this work, we utilize a modified DenseNet architecture as proposed by \cite{Huang2017dcc}, where the outputs of earlier layers are concatenated to all the consecutive layers. We simplify the structure such that the model only contains skip connections from the input layer to all consecutive layers. We call this architecture InputSkip, which has reduced complexity compared to DenseNet:
\begin{equation} \label{eq:inputskip}
\begin{aligned}
    \mathbf{z}_{1} &= \boldsymbol{\sigma}_1(\mathbf{W}_1\mathbf{z}_{0} + \mathbf{b}_1) \\
    \mathbf{z}_{i} &= \boldsymbol{\sigma}_i\left(\mathbf{W}_i\begin{bmatrix}\mathbf{z}_{i-1} \\ \mathbf{z}_{0}\end{bmatrix} + \mathbf{b}_i\right) , i>1
\end{aligned}
\end{equation}
The output of each layer (excl. the first) becomes a sum of a linear and a nonlinear transformation of the initial input $\mathbf{x}$. Hence, the skip connections from the input layer to consecutive layers facilitate the reuse of the input features for modeling different linear and nonlinear relationships more independently. The InputSkip architecture with four hidden layers is illustrated in Figure~\ref{fig:inputSkip}.

\section{Method and setup}
In this section, we present all the details of data generation, its preprocessing, and the methods required to reproduce the work. The steps can be briefly summarized as follows:
\begin{itemize}
    \item Use \eqref{eq:alu_equations} with random initial conditions to generate 140 trajectories with 5000 timesteps each. Set aside 40 for training and 100 for testing. Construct three datasets by selecting 10,20, and 40 trajectories, respectively.
    \item For each model class and dataset, train ten instances on the training data.
    \item Repeat all experiments with $\ell_1$ regularization, see loss function in \eqref{eq:l1}.
    \item Use trained models to generate predicted trajectories along the test set and compare them to the 100 test trajectories.
\end{itemize}



\subsection{Data generation}
\label{sec:data-generation}

The state trajectories used in the test and training sets were generated by integrating Equation~\eqref{eq:alu_equations} with the numerical RK4 integration scheme with a fixed timestep $\timestep = \SI{10}{\second}$ on the interval $[0, 5000\timestep]$. 
The initial conditions were sampled uniformly from the intervals shown in Table~\ref{table:init_conditions_aluminum} to generate 140 unique trajectories. 
A total of 40 trajectories were set aside for training and 100 of the trajectories as a test set. 
The 40 training trajectories were used to create three datasets of varying sizes (small, medium, large), namely 10, 20, and 40 trajectories, containing 50000, 100000, and 200000 individual data points.

Equation \eqref{eq:alu_equations} also depends on the input signal $\xinput$. 
In practice, this is given by a deterministic control policy $\xinput = \boldsymbol{\pi}(\state)$ that stabilizes the system and keeps the state $\state$ within some region of the state space that is suitable for safe operation.
We found that this was insufficient to successfully train our models because the controlled trajectories showed minimal variation after some time, despite having different initial conditions. 
This lack of diversity in the dataset resulted in models that could not generalize to unseen states, which frequently arose during evaluation.
To inject more variety into the data and sample states $\state$ outside of the standard operational area, we used a stochastic controller
$$\boldsymbol{\pi}_s(\state) = \boldsymbol{\pi}(\state) + \mathbf{r}(t)$$ 
that introduced random perturbations $\mathbf{r}(t)$ to the input.
These perturbations were sampled using the \ac{APRBS} method proposed by \cite{WINTER2018802}.

In system identification, it is typical to optimize the model to estimate the function $\mathbf{\dot{x} = f(x,u)}$. However, this is not feasible for \eqref{eq:alu_equations} because the inputs $\xinput$ are not differentiable. Instead, the trajectories are discretized using the forward Euler difference:
\begin{equation}\label{eq:forward_euler}
    \mathbf{y}_k = \frac{\state_{k+1} - \state_{k}}{\timestep}    
\end{equation}
The datasets are then constructed as sets of the pairs $([\state_k,\xinput_k], \mathbf{y}_k)$.
In practice, measurements will be noisy and the state trajectories must be estimated using a filtering method, e.g., moving horizon estimation.


\begin{table}
    \centering
    \caption{Initial conditions intervals for $\state$}
    \label{table:init_conditions_aluminum}
    \begin{tabular}{l|l}
    \hline
    Variable &  Initial condition interval \\ \hline
    $x_1$ & $[2060,\;4460]$\\
    $c_{x_2}$ & $[0.02,\; 0.05]$ \\
    $c_{x_3}$ & $[0.09,\; 0.13]$  \\
    $x_4$& $[11500,\; 16000]$  \\
    $x_5$ & $[9550,\; 10600]$ \\
    $x_6$ & $[940,\; 990]$ \\
    $x_7$ & $[790,\; 850]$ \\
    $x_8$ & $[555,\; 610]$ \\
    \hline
    \end{tabular}
\end{table}

\subsection{Model architectures}
Two different architectures are evaluated in this case study: a standard feed-forward Multi-Layer Perceptrons (MLP) referred to as PlainNet and the modified MLP with concatenated skip-connections from the input layer called InputSkip, see Fig.~\ref{fig:inputSkip} for illustration. Moreover, both structures are trained with and without the sparsity promoting $\ell_1$ regularization, yielding four model structures: PlainDense, PlainSparse, InputSkipDense, and InputSkipSparse. The input layer of each of the models is the concatenation of the measured state $\mathbf{x}_k \in \mathbb{R}^8$, or the estimated state $\hat{\mathbf{x}}_k\in \mathbb{R}^8$ at timestep $k$, and the control input vector $\mathbf{u}_k\in \mathbb{R}^8$ at timestep $k$, yielding a vector $\mathbf{z}_0 = \{\mathbf{x}_k, \; \mathbf{u}_k\} \in  \mathbb{R}^{13}$. Each of the structures has four hidden layers, and each of the layers has 25 neurons. In addition, the input vector $\mathbf{z_0}$ is concatenated to each of the hidden layers in the InputSkip structures, such that each of the hidden layers in the InputSkip structures has $25 + 13 = 38$ states. All models output an estimate of the time derivative of the state variables at timestep $k$ $\Dot{\mathbf{x}}_k \in \mathbb{R}^8$. Each model class's sparse and dense structures start with the same architecture before training, but for sparse structures, many of the neurons and weights are zeroed out by the $\ell_1$ regularization term. Since InputSkip has more states in the hidden layers than Plain structures due to the input vector being concatenated to each layer's output, it is reasonable to ask whether the Plain structures should have more neurons in the hidden layers. This is tested, and it turns out that this does not benefit the structures regarding the evaluation measures.

\subsection{Training setup}
\label{subsec:training}

The models are trained by minimizing the following loss function using stochastic gradient descent:
\begin{equation}
    \label{eq:l1}
    \mathbf{J}_{\theta} = \frac{1}{|\mathcal{B}|}\sum_{i\in\mathcal{B}} (\mathbf{\mathbf{y}_i}-\mathbf{\net(\state_i, \xinput_i}))^2 + \lambda \sum_{j=1}^{L}|\mathbf{W}_j|
\end{equation}
where the \textit{batch} $\mathcal{B}$ is a set of indices corresponding to a random subset of examples from the data, $L$ is the number of layers of the \ac{NN}, and $\lambda$ is the regularization parameter. This loss function is the sum of the \ac{MSE} of the model $\net$ with respect to the regression variables $\mathbf{y}$, and the $\ell_1$ norm of the connection weight matrices $\mathbf{W}_i$ in all layers. We used a batch size of $|\mathcal{B}| = 128$. We used the popular ADAM solver proposed by \cite{kingma_adam_2014} with default parameters to minimize \eqref{eq:l1}. The dense model structures PlainDense and InputSkipDense were trained with $\lambda=0$, and the sparse model structures PlainSparse and InputSkipSparse were trained with $\lambda=10^{-4}$.

\subsection{Evaluation of model accuracy}
\label{subsec:performancemetrics}

Starting from a given initial condition $\mathbf{x}(t_0)$, the model $\net(\state, \xinput)$ is used to generate an estimated trajectory using the recurrence:
\begin{equation}
\label{eq:euler_forecast}
\hat{\state}_{k+1} = \hat{\state}_{k} + \timestep\,\net(\hat{\state}_{k}, \xinput_k)
\end{equation}
where $\hat{\state}_{0} = {\state}_{0}$.
Applying multi-step or higher order Runge-Kutta methods to \ac{NN} models is possible. However, these methods require multiple evaluations of the model per timestep, which increases the computation and memory needed to perform automatic differentiation. 
The forward Euler method is preferred, as it only evaluates the model once per timestep.
In the approach outlined here, this does not incur significant discretization errors, as \eqref{eq:euler_forecast} effectively reverses the discretization step in \eqref{eq:forward_euler}.
The input signal $\xinput_k$ is sampled directly from the test trajectory.
Borrowing a term from the field of time-series analysis, this is referred to as a \textit{rolling forecast}. 
To evaluate the accuracy of a model over multiple trajectories, we define the Average Normalized Rolling Forecast Mean Squared Error (AN-RFMSE):
\begin{equation}
    \textrm{AN-RFMSE} = \frac{1}{p}\sum_{i=1}^p\frac{1}{n}\sum_{j=1}^n\left(\frac{\hat{x}_i(t_j) - x_i(t_j)}{\std(x_i)}\right)^2,
    \label{eq:AN-RFMSE}
\end{equation}
where $\hat{x}_i(t_j)$ is the model estimate of the simulated state variable $x_i$ at time step $t_j$, $\std(x_i)$ is the standard deviation of variable $x_i$ in the training set $\mathcal{S}_{train}$, $p=8$ is the number of state variables and $n$ is the number of time steps being averaged over.

\subsection{Evaluation of model stability}
\label{subsec:blowup}
A symptom of model instability is that its predictions can \textit{blow up}, characterized by a rapid (often exponential) increase in prediction error. More precisely, a blow-up is said to occur when all system states' normalized mean absolute error exceeds three (this corresponds to standard deviations):
\begin{equation}
    \max_{j<n} \left[\frac{1}{p}\sum_{i=1}^p\left(\frac{\left\vert\hat{x}_i(t_j) - x_i(t_j)\right\vert}{\std(x_i)}\right)\right] > 3
    \label{eq:blowup-detection}
\end{equation}
where $p=8$ is again the number of state variables and $n$ is the number of time steps to consider.
Equation~\eqref{eq:blowup-detection} is conservative. However, this does not lead to a significant underestimation of the number of blow-ups. This is because once a model starts to drift rapidly, it quickly exceeds the threshold.

\section{Results and discussions}
\label{sec:resultsanddiscussions}

\begin{figure}
    \centering
    \begin{subfigure}[t]{\linewidth}
    \includegraphics[width=\linewidth]{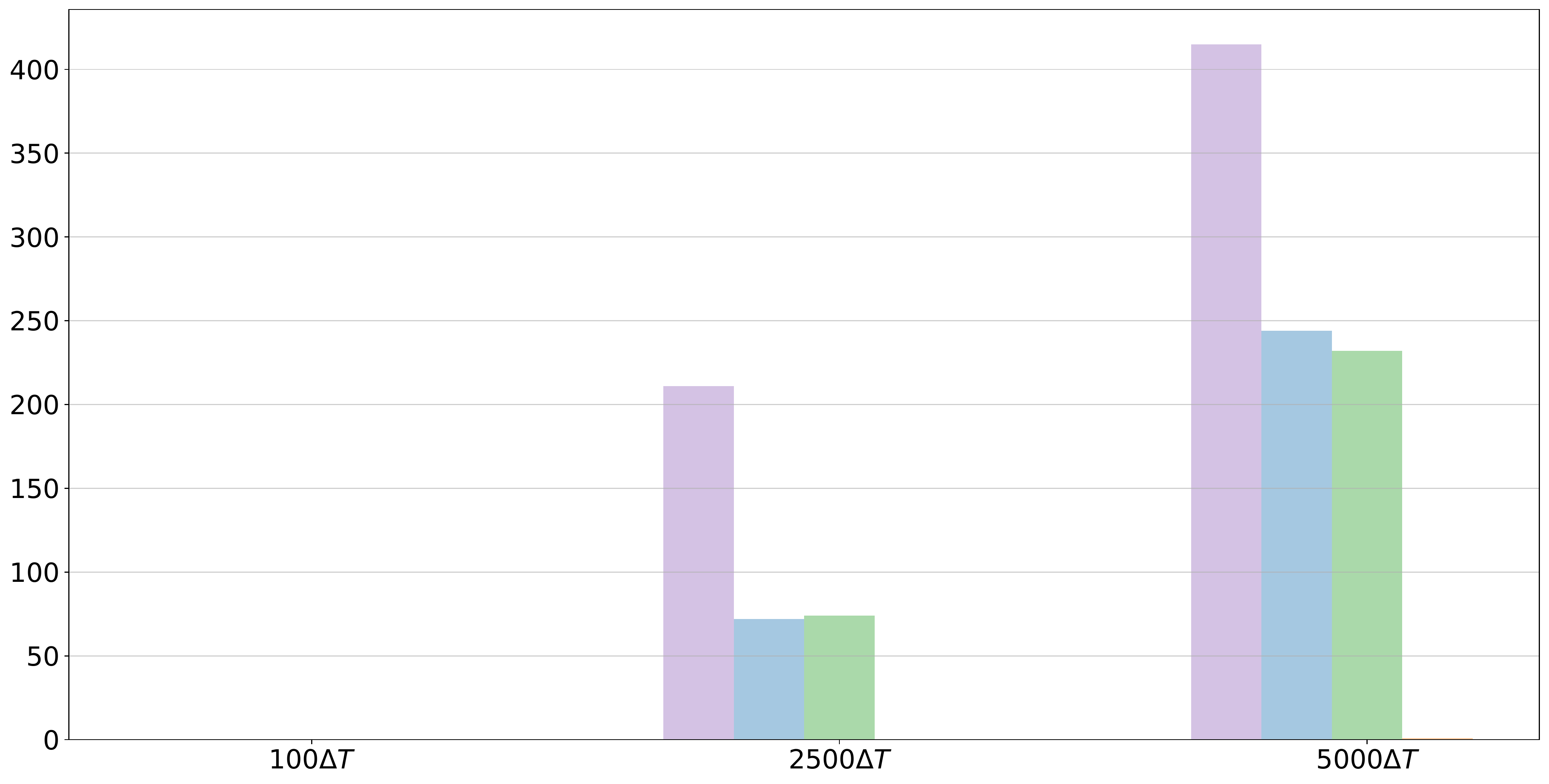}
    \caption{Trained on smallest dataset with 50000 datapoints}
    \label{subfig:div_plot_dSet0}
    \end{subfigure}
    \begin{subfigure}[t]{\linewidth}
    \includegraphics[width=\linewidth]{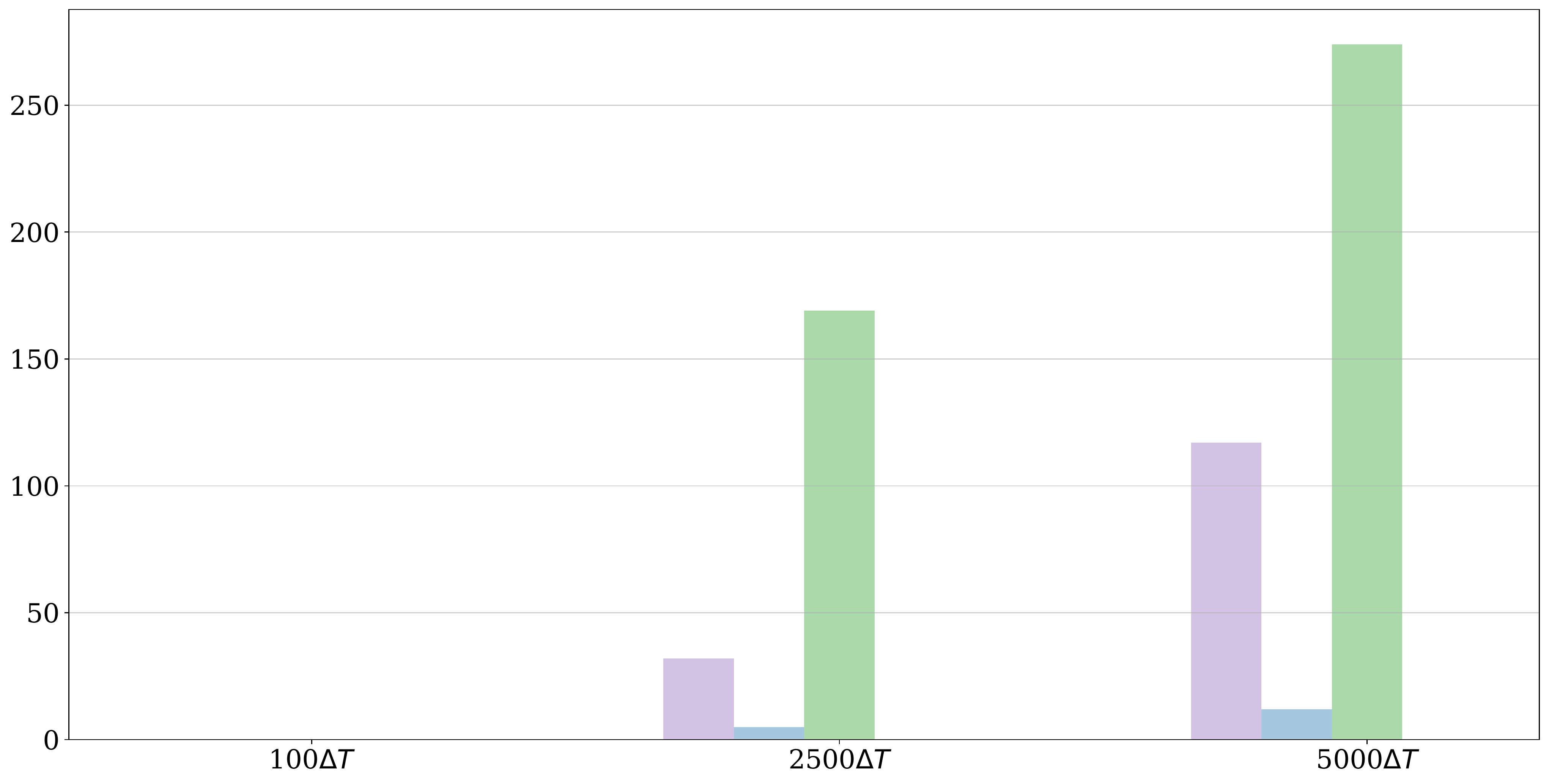}
    \caption{Trained on medium sized dataset with 100000 datapoints}
    \label{subfig:div_plot_dSet1}
    \end{subfigure}
    \begin{subfigure}[t]{\linewidth}
    \includegraphics[width=\linewidth]{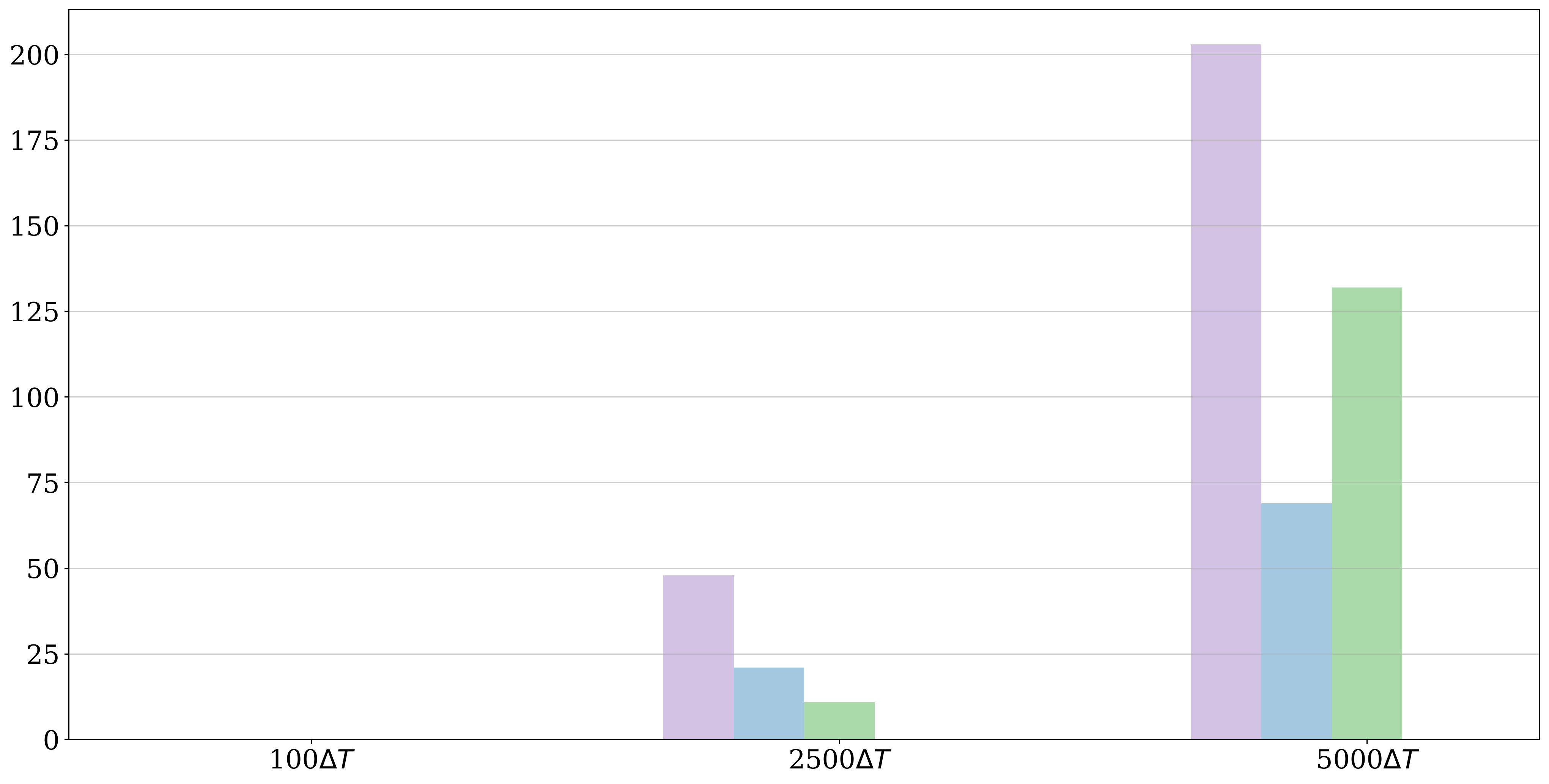}
    \caption{Trained on largest dataset with 200000 datapoints}
    \label{subfig:div_plot_dSet2}
    \end{subfigure}
    \begin{adjustbox}{max width=\linewidth}
        \begin{tikzpicture}
            \begin{customlegend}[legend columns=4,legend style={draw=none, column sep=1ex},legend entries={ PlainDense,PlainSparse,InputSkipDense,InputSkipSparse}]
                \addlegendimage{Tpurple!50, fill=Tpurple!50,area legend}
                \addlegendimage{Tblue!50, fill=Tblue!50,area legend}
                \addlegendimage{Tgreen!50, fill=Tgreen!50,area legend}
                \addlegendimage{Torange!50, fill=Torange!50,area legend}
            \end{customlegend}
        \end{tikzpicture}
        \end{adjustbox}
    \caption{Divergence plot: Number of trajectories that blow-up over different time horizons according to the measure defined in Equation~\eqref{eq:blowup-detection}. The total number of trajectories is 1000, so the values can be read as a permille. }
    \label{fig:Divergence_plot}
\end{figure}

\begin{figure}
    \centering
    \begin{subfigure}[t]{\linewidth}
        \includegraphics[width=\linewidth]{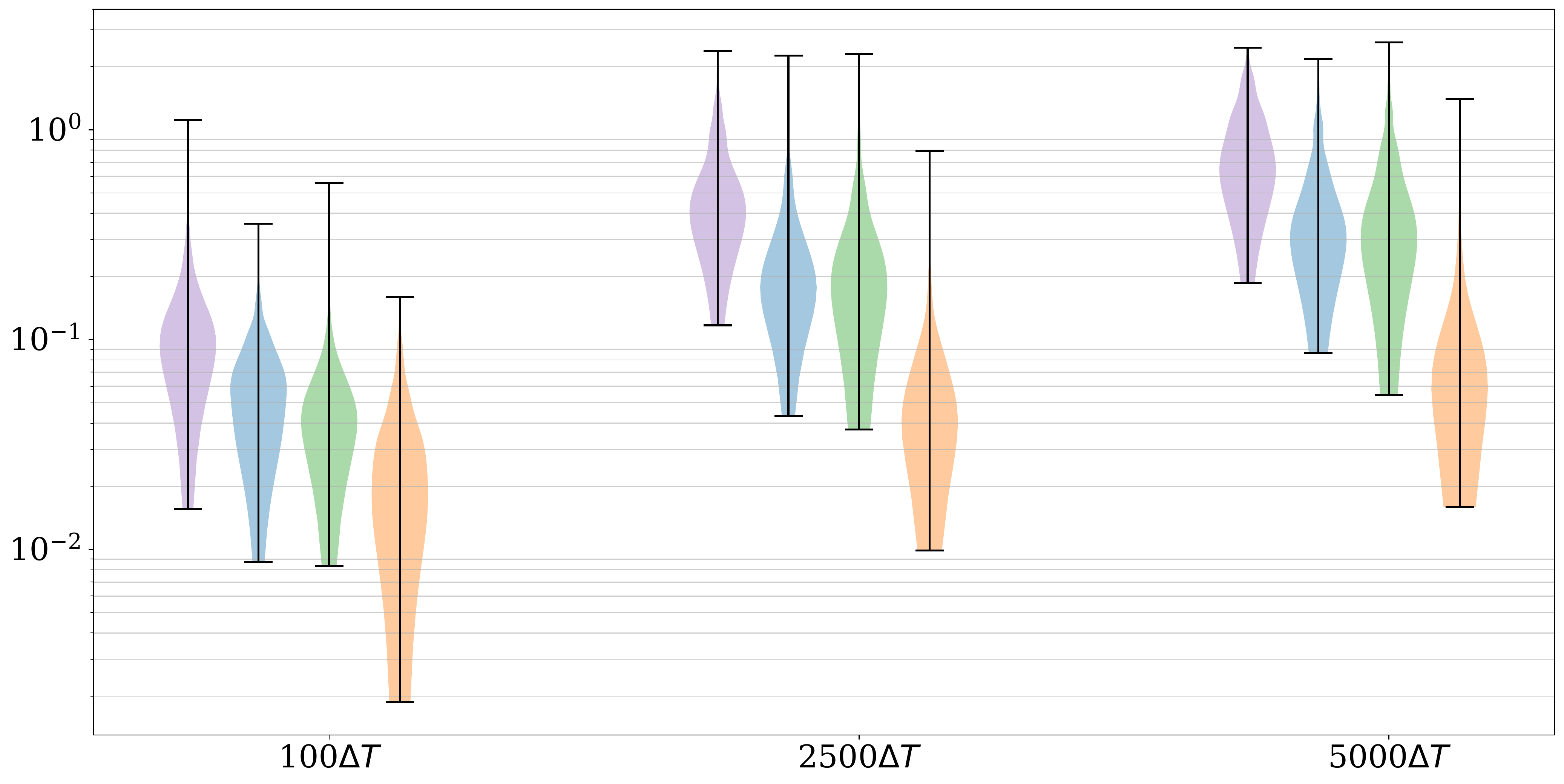}
        \caption{Trained on smallest dataset with 50000 datapoints}
        \label{subfig:violin_plot_dSet0}
    \end{subfigure}
    \begin{subfigure}[t]{\linewidth}
        \includegraphics[width=\linewidth]{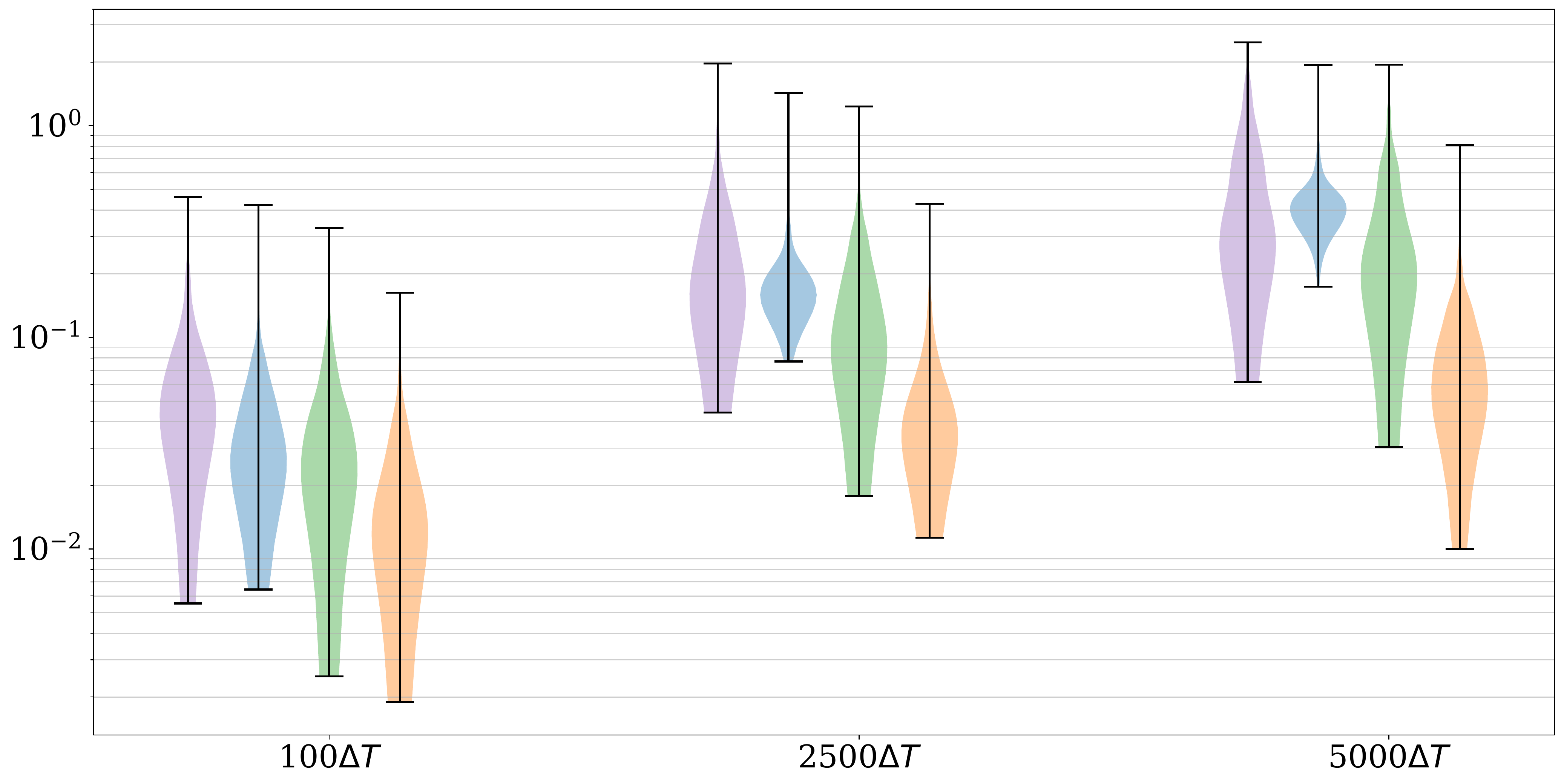}
        \caption{Trained on medium sized dataset with 100000 datapoints}
        \label{subfig:violin_plot_dSet1}
    \end{subfigure}
    \begin{subfigure}[t]{\linewidth}
        \includegraphics[width=\linewidth]{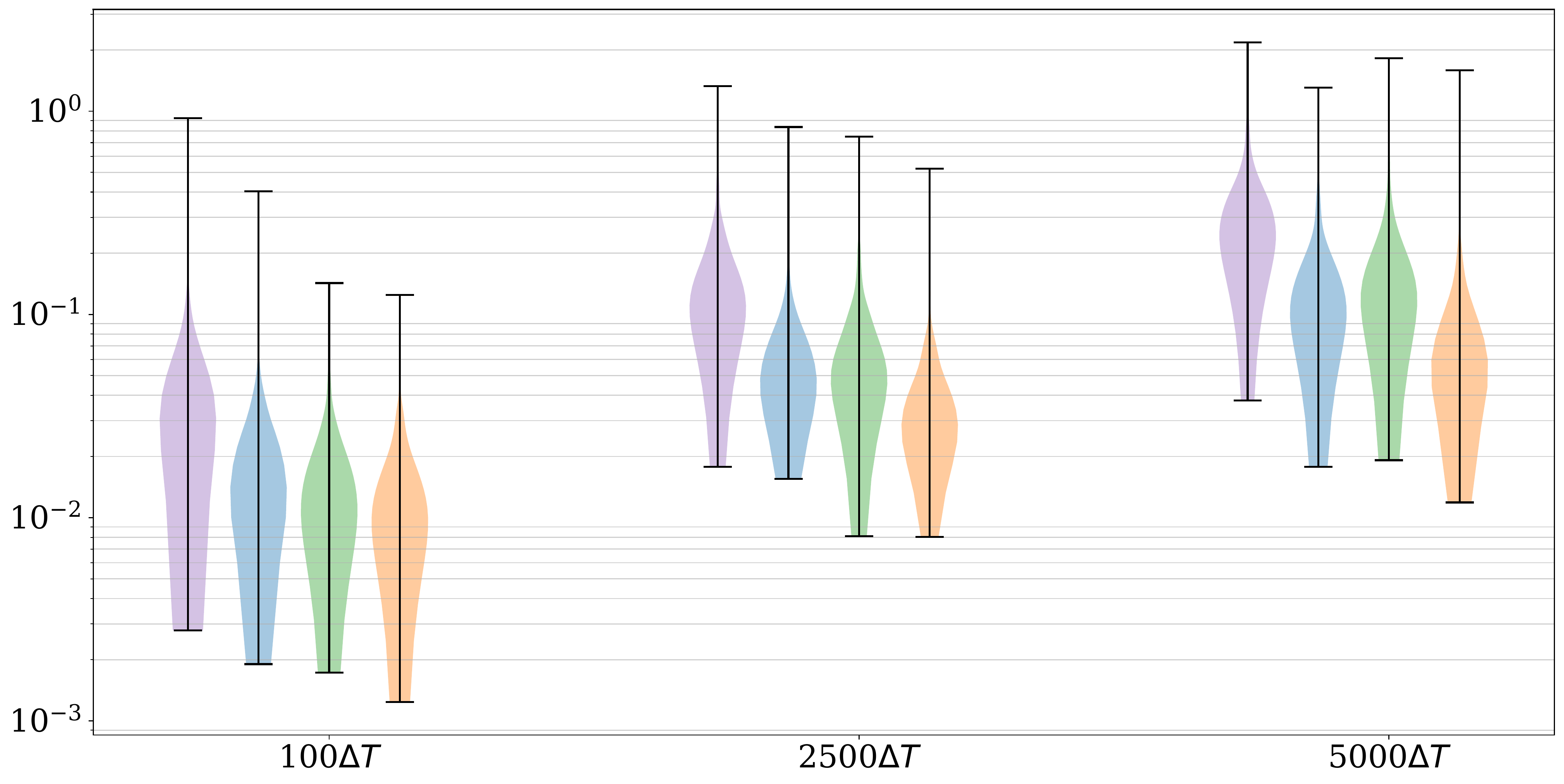}
        \caption{Trained on largest dataset with 200000 datapoints}
        \label{subfig:violin_plot_dSet2}
    \end{subfigure}
    \begin{adjustbox}{max width=\linewidth}
        \begin{tikzpicture}
            \begin{customlegend}[legend columns=4,legend style={draw=none, column sep=1ex},legend entries={ PlainDense,PlainSparse,InputSkipDense,InputSkipSparse}]
                \addlegendimage{Tpurple!50, fill=Tpurple!50,area legend}
                \addlegendimage{Tblue!50, fill=Tblue!50,area legend}
                \addlegendimage{Tgreen!50, fill=Tgreen!50,area legend}
                \addlegendimage{Torange!50, fill=Torange!50,area legend}
            \end{customlegend}
        \end{tikzpicture}
        \end{adjustbox}
    \caption{Model accuracy is expressed in terms of AN-RFMSE over different horizons. AN-RFMSE is defined for a single model forecast of a single trajectory in \eqref{eq:AN-RFMSE}. Ten models of each of the model types (PlainDense, PlainSparse, InputSkipDense, InputSkipSparse) are trained on the 50000 data points in Figure~\ref{subfig:violin_plot_dSet0}, 100000 data points in Figure~\ref{subfig:violin_plot_dSet1}, and 200000 data points in Figure~\ref{subfig:violin_plot_dSet2}. The model estimates that blow up (see Figure~\ref{fig:Divergence_plot}) are excluded. The plot shows that sparse models with skip connections (InputSkipSparse) are consistently more accurate than sparse and dense models without skip connections.}
    \label{fig:Violin_plot}
\end{figure}

This section reports empirical results for the model accuracy and stability of the different model classes (PlainDense, PlainSparse, InputSkipDense, InputSkipSparse).
A Monte Carlo analysis is performed by training ten instances of each model class and evaluating these on the test set consisting of 100 trajectories, where each trajectory has a length of 5000 timesteps. The test set is generated as described in Section~\ref{sec:data-generation}, using the simulation model in Equation~\eqref{eq:alu_equations}. The evaluation procedure follows; the models are given initial values for each state variable trajectory in the test set. Then, the models forecast state values at the consecutive time steps for each test set trajectory as described in \eqref{eq:euler_forecast} without feedback from measurements of the state variables. The resulting forecasts are evaluated according to the prediction accuracy measure in \eqref{eq:AN-RFMSE} and the forecast stability measure in \eqref{eq:blowup-detection}. The prediction accuracy of different model classes is reported in Fig.~\ref{fig:Violin_plot}, and the empirical model stability results are reported in Fig.~\ref{fig:Divergence_plot}. Accuracy and stability are reported for different forecasting horizons. We repeat the experiments for all model classes trained on three different dataset sizes to study the data efficiency of the models.

Fig.~\ref{fig:Divergence_plot} presents the total number of blow-ups recorded within each model class after $100\timestep$, $2000\timestep$, and $5000\timestep$ (short, medium, and long term respectively). 
For simplicity, blow-ups were detected by thresholding the computed variance of a predicted trajectory and manually inspected.
It is clear that for short time horizons, all the models exhibit robust behavior independently of the size of the training datasets. However, for medium and long time horizons, PlainDense, PlainSparse, and InputSkipDense architectures exhibit a significant number of blow-ups and, therefore, instability. Figs.~\ref{subfig:div_plot_dSet0} - \ref{subfig:div_plot_dSet2} show that PlainDense is generally the most unstable, with up to 41\% of all trajectories resulting in a blow-up. For the smallest amount of training data (see Fig.~\ref{subfig:div_plot_dSet0}) PlainSparse and InputSkipDense have similar blow-up frequencies. The PlainSparse architecture shows significantly better stability for larger datasets than both PlainDense and InputSkipDense. InputSkipDense and PlainDense both show better stability with increasing training data regarding fewer blow-ups. However, both these dense models still suffer from high blow-up rates.

In comparison, almost no blow-ups are recorded using the InputSkipSparse architecture, even for the small training dataset. In Figure \ref{fig:Divergence_plot}, the orange bars corresponding to the blow-up frequency of InputSkipSparse models are not visible for any training sets due to the significantly lower number of blow-ups. For InputSkipSparse models trained on the smallest dataset, only 3 out of 1000 possible blow-ups were reported for the longest horizon. Apart from that, no blow-ups were reported for the InputSkipSparse models. 
Only a few blow-ups were recorded after $5000\timestep$ in the medium term.

Fig.~\ref{fig:Violin_plot} presents a violin plot of the accuracy of each model class, expressed in terms of AN-RFMSE over different time horizons. 
A larger width of the violin indicates a higher density of that given RFMSE value, while the error bars show the minimum and maximum recorded RFMSE values. 
The model estimates that blew up (see Fig.~\ref{fig:Divergence_plot}) are excluded as outliers.
In this way, the generalization performance of the models is estimated only within their regions of stability.
A potential pitfall of excluding these outliers is that model classes that blow up often have their worst scores removed, thus biasing the distribution towards lower scores.
Despite this, low accuracy appears to correlate with a high blow-up rate. The InputSkipSparse architecture is consistently more accurate (up to an order of magnitude) than the others in the long term.

\begin{figure*}
    \centering
    \begin{subfigure}[t]{0.475\linewidth}
    \includegraphics[width=\linewidth]{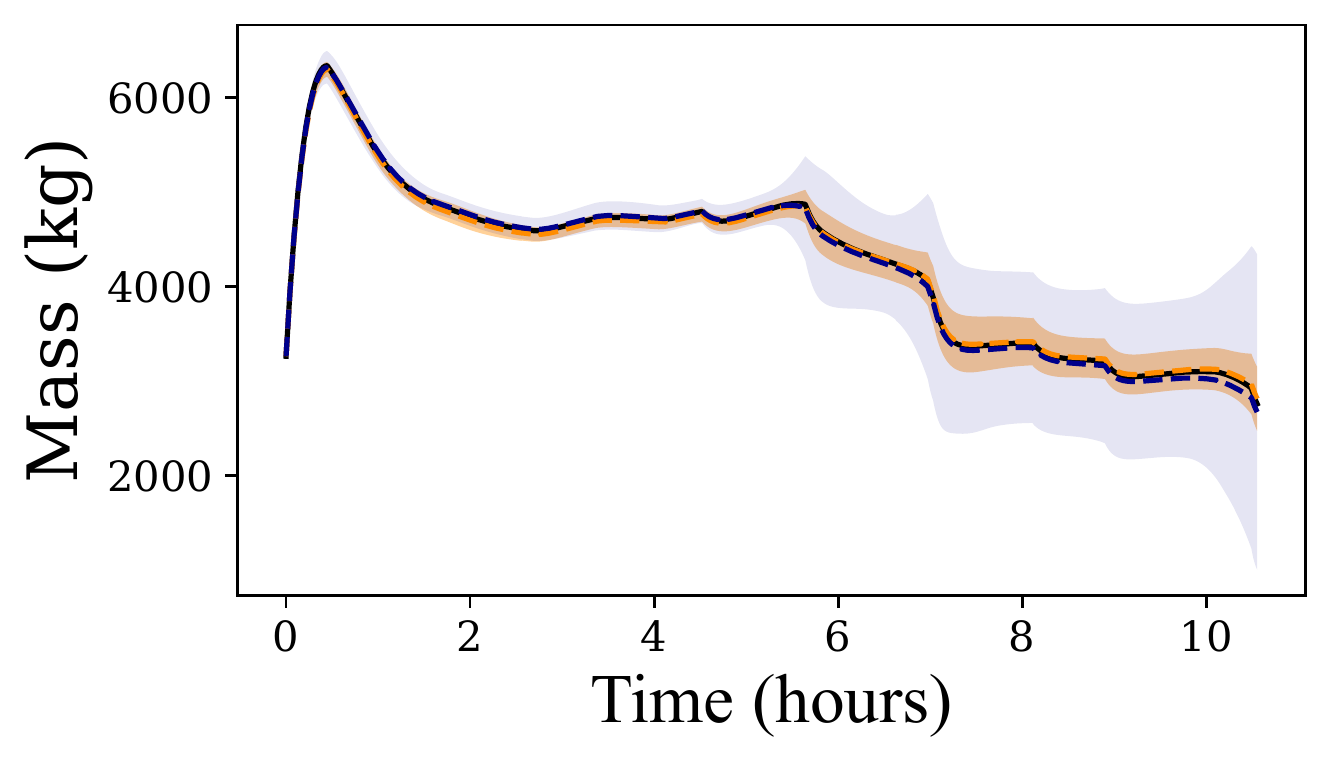}
    \label{subfig:RF_x1_dSet0}
    \caption{Side ledge mass $x_1$}
    \end{subfigure}
    \begin{subfigure}[t]{0.475\linewidth}
    \includegraphics[width=\linewidth]{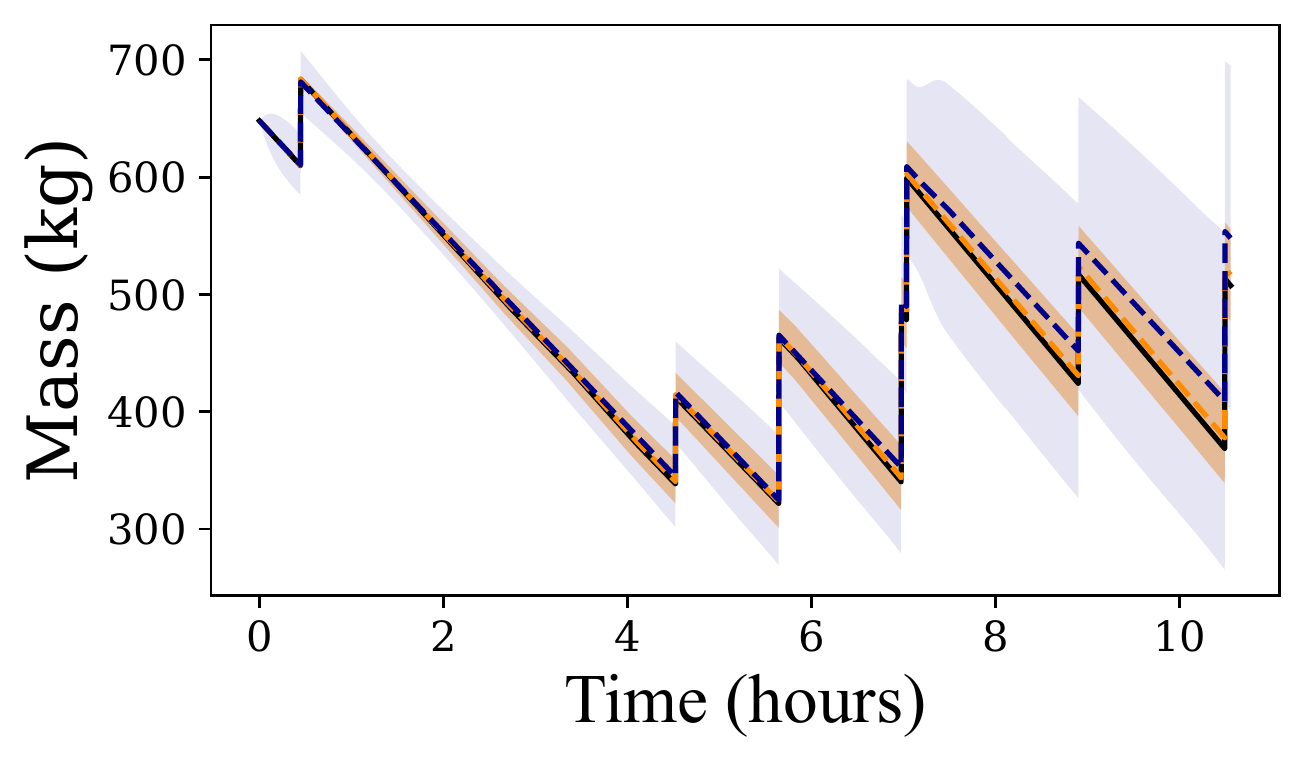}
    \caption{Alumina mass $x_2$}
    \label{subfig:RF_x2_dSet0}
    \end{subfigure}
    \begin{subfigure}[t]{0.475\linewidth}
    \includegraphics[width=\linewidth]{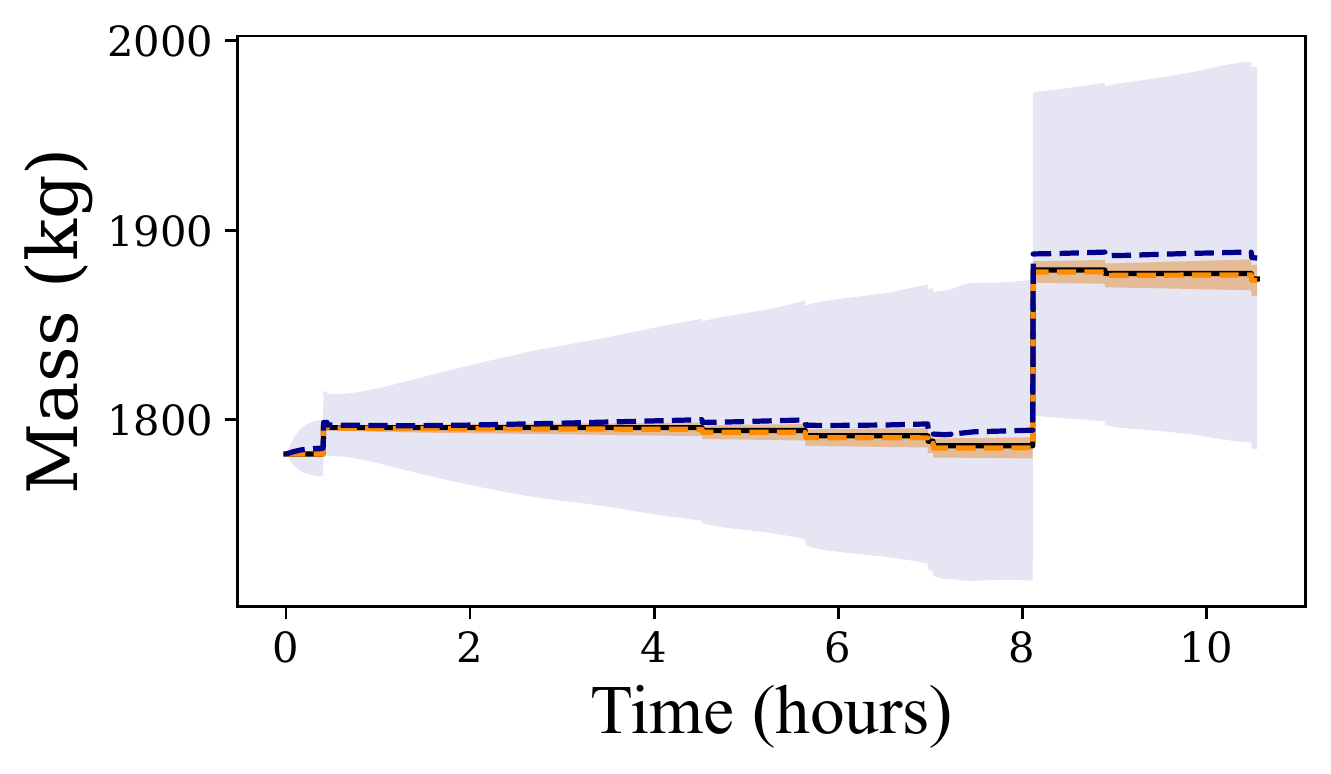}
    \caption{Aluminum fluoride $x_3$}
    \label{subfig:RF_x3_dSet0}
    \end{subfigure}
    \begin{subfigure}[t]{0.475\linewidth}
    \includegraphics[width=\linewidth]{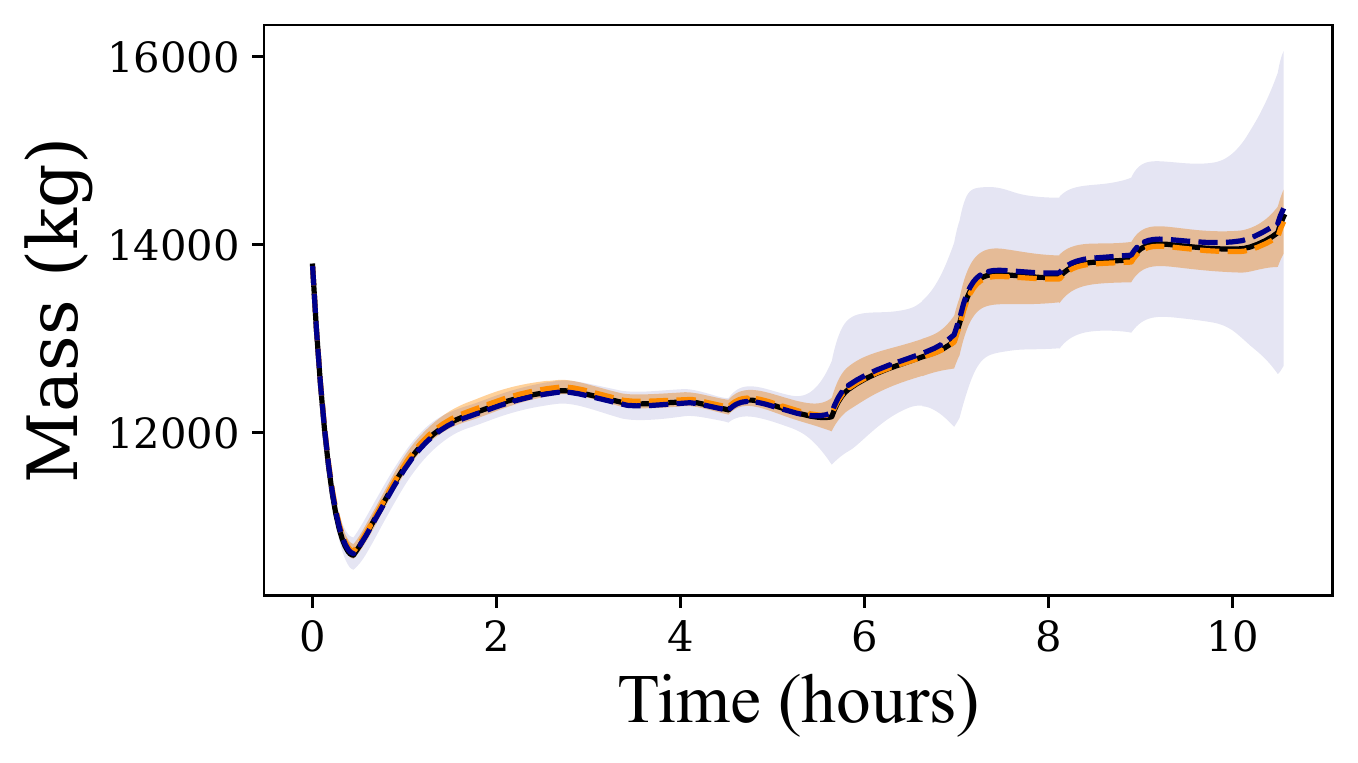}
    \caption{Molten cryolite $x_4$}
    \label{subfig:RF_x4_dSet0}
    \end{subfigure}
    \begin{subfigure}[t]{0.475\linewidth}
    \includegraphics[width=\linewidth]{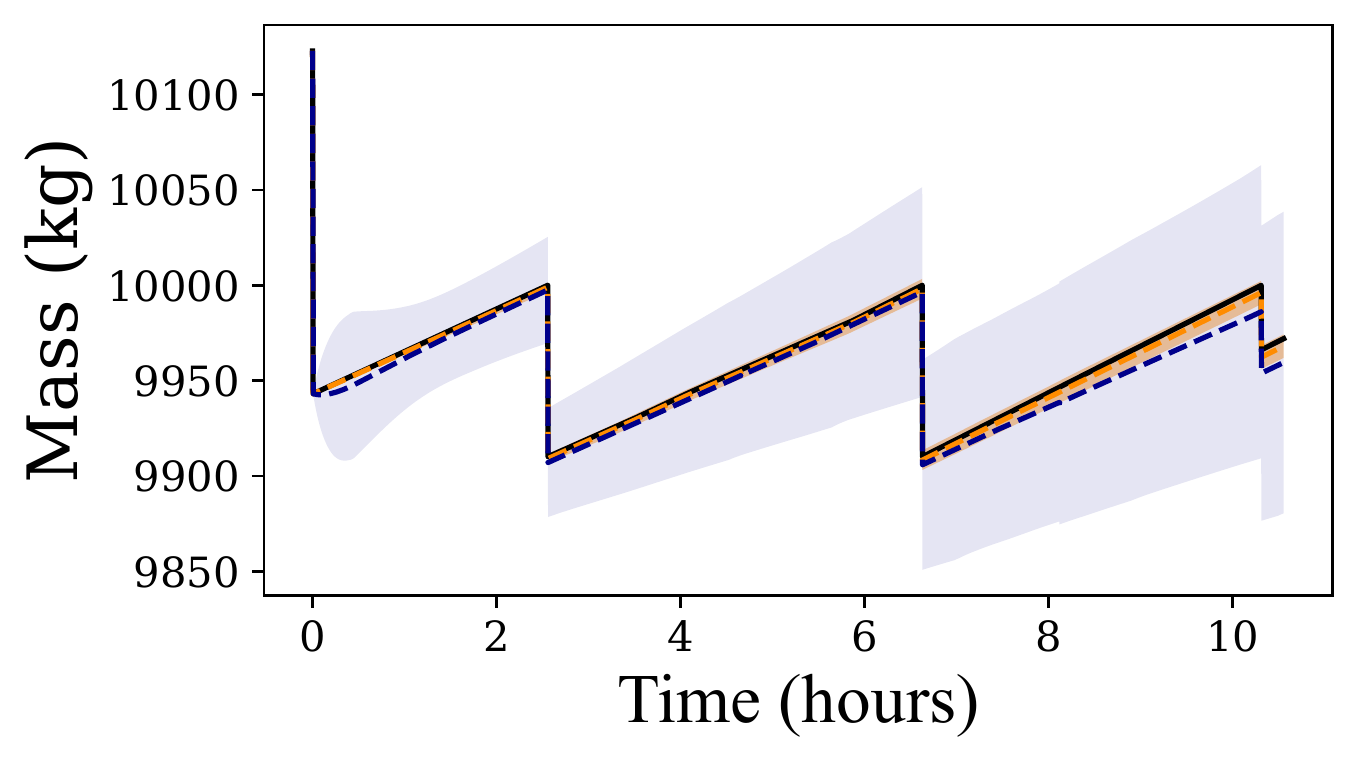}
    \caption{Produced aluminum $x_5$}
    \label{subfig:RF_x5_dSet0}
    \end{subfigure}
    \begin{subfigure}[t]{0.475\linewidth}
    \includegraphics[width=\linewidth]{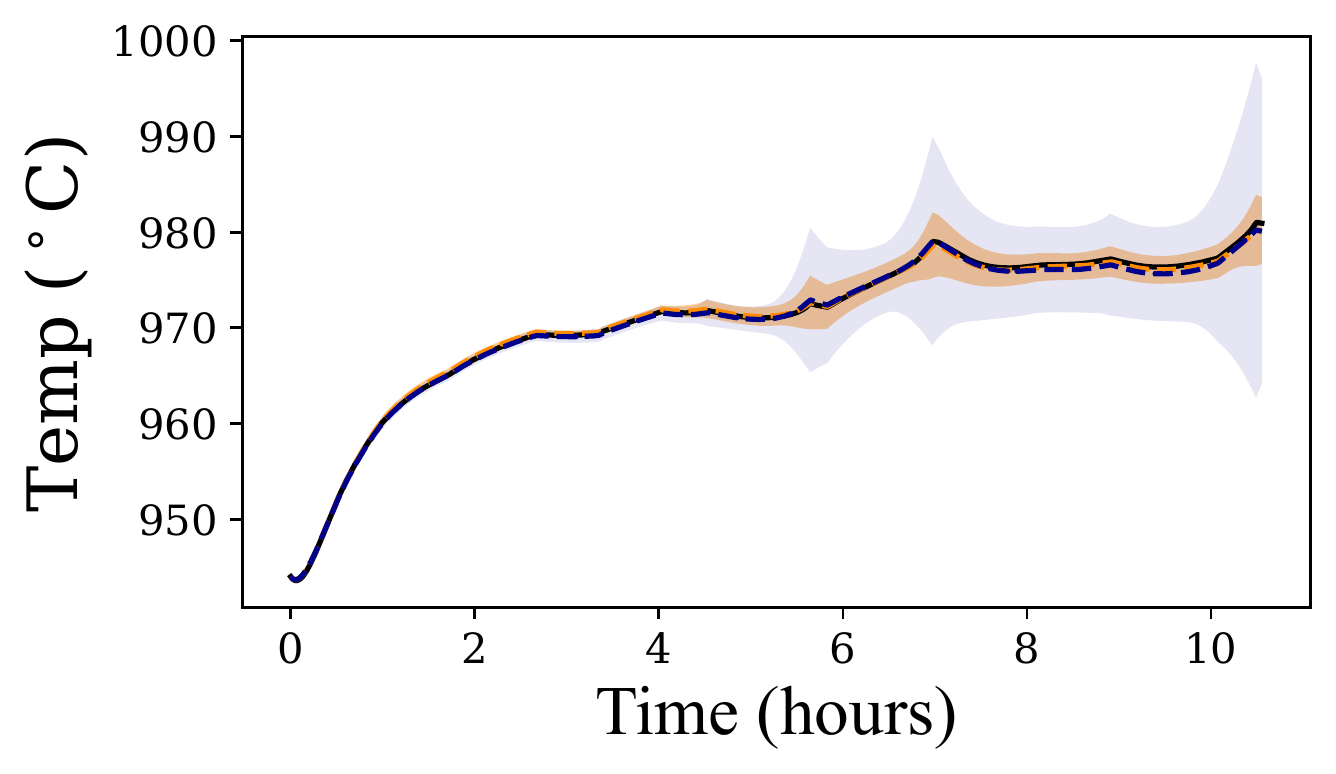}
    \caption{Bath temperature $x_6$}
    \label{subfig:RF_x6_dSet0}
    \end{subfigure}
    \begin{subfigure}[t]{0.475\linewidth}
    \includegraphics[width=\linewidth]{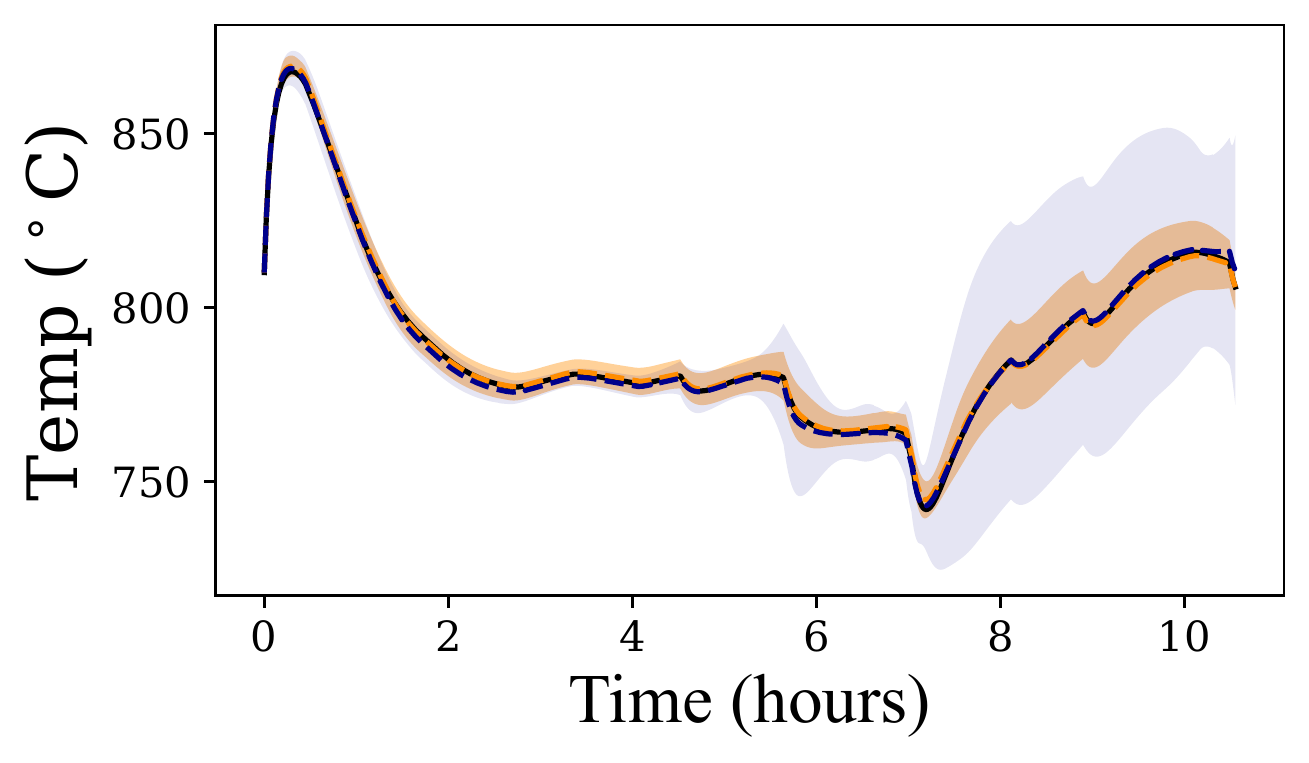}
    \caption{Side ledge temperature $x_7$}
    \label{subfig:RF_x7_dSet0}
    \end{subfigure}
    \begin{subfigure}[t]{0.475\linewidth}
    \includegraphics[width=\linewidth]{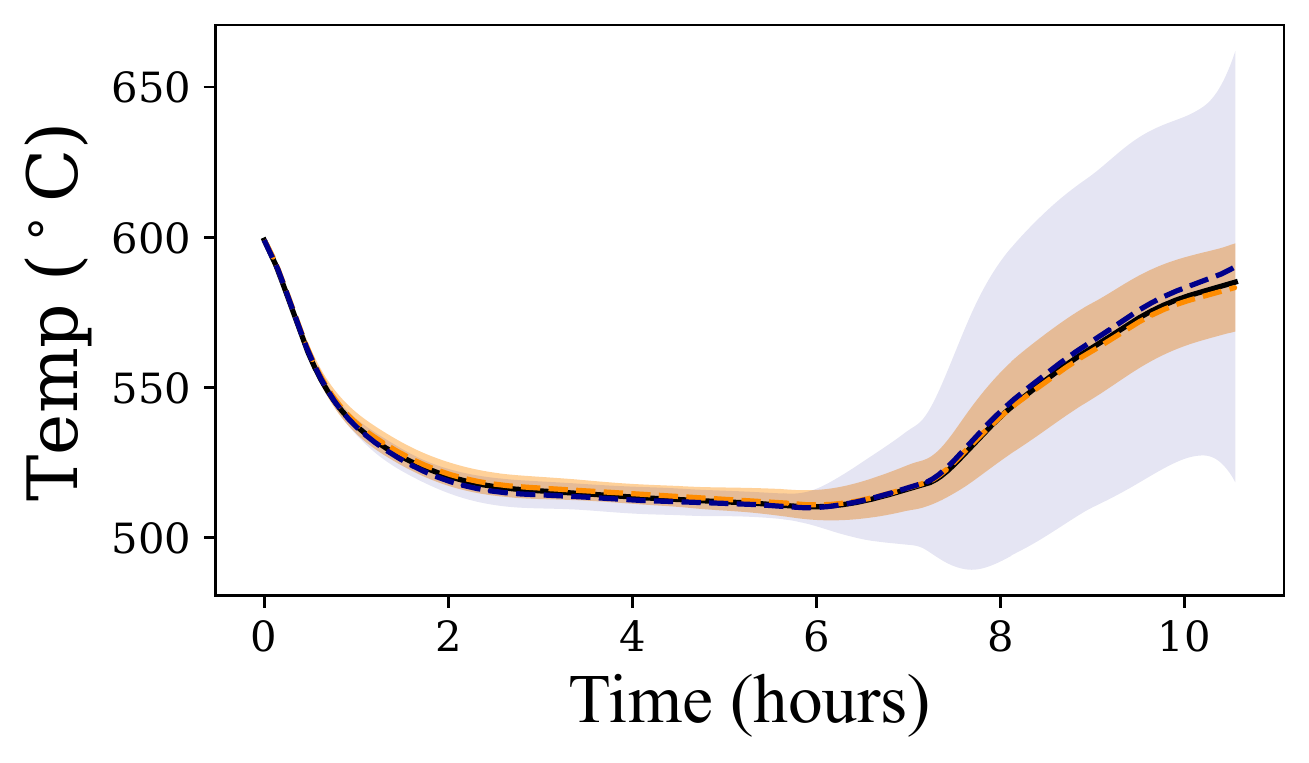}
    \caption{Side wall temperature $x_8$}
    \label{subfig:RF_x8_dSet0}
    \end{subfigure}
            \begin{adjustbox}{max width=1\linewidth}
        \begin{tikzpicture}
            \begin{customlegend}[legend columns=6,legend style={draw=none, column sep=2ex},legend entries={Truth,InputSkipSparse,PlainSparse,$99.7\%$ conf. PlainSparse,$99.7\%$ conf. InputSkipSparse}]
                \addlegendimage{black,thick, sharp plot}
                \addlegendimage{Torange,thick, dashed,sharp plot}
                \addlegendimage{Tblue,thick, dashed,sharp plot}
                \addlegendimage{Tpurple!30, fill=Tpurple!20, area legend}
                \addlegendimage{Torange!40, fill=Torange!40, area legend}
            \end{customlegend}
        \end{tikzpicture}
    \end{adjustbox}
    \caption{Rolling forecast of a representative trajectory from the test set (100 trajectories total)}
    \label{fig:RF_X}
\end{figure*}
Fig.~\ref{fig:RF_X} shows the effect of InputSkipSparse compared to PlainSparse for a single representative test-set trajectory and is not meant to the significance of the results. The significance of the results can be found in Fig.~\ref{fig:Violin_plot} and Fig.~\ref{fig:Divergence_plot}, which show results for the entire test set.

\section{Conclusion and future work}
\label{sec:conclusions}
This work compared the performance of two different model structures trained with and without sparsity promoting $\ell_1$ regularization. The two model types are standard MLP and a more specialized architecture that includes skip connections from the input layer to all consecutive layers, yielding four different model structures: PlainDense, PlainSparse, InputSkipDense, and InputSkipSparse. The main conclusions of the article are as follows:
\begin{itemize}
    \item \ac{NN}s with skip connections are more stable for predictions over long time horizons compared to standard MLPs. Furthermore, the accuracy of \ac{NN}s with skip connections is consistently higher for all forecasting horizons. 
    \item The application of sparsity-promoting $\ell_1$ regularization significantly improves the stability of the standard MLP and InputSkip architectures. This improvement was more apparent for models with the InputSkip architecture.
    \item The InputSkipSparse showed satisfactory stability characteristics even when the amount of training data was restricted, suggesting that this architecture is more suitable for system identification tasks than the standard MLP structure.
\end{itemize}
The case study shows that both sparsity-promoting regularization and skip connections can result in more stable \ac{NN} models for system identification tasks while requiring fewer data and improving their multi-step generalization for both short, medium, and long prediction horizons. 
Despite the encouraging performance of the sparse-skip networks, it is yet to be determined if the benefits also extend to the case of noisy measurements.

This case study also has relevance beyond the current case study.
In more realistic situations, we often have a partial understanding of the system we wish to model (see \eqref{eq:alu_equations}) and only wish to use data-driven methods to correct a \ac{PBM} when it disagrees with the observations (e.g., due to a faulty assumption).
As shown in \cite{robinson2022anc}, combining \ac{PBM}s and data-driven methods in this way also has the potential to inject instability into the system.
Finding new ways to improve or guarantee out-of-sample behavior for data-driven methods is therefore paramount to improving such systems' safety.

\bibliography{ref.bib}             

\begin{thebibliography}{10}

\bibitem{allenzhu_convergence_2019}
Zeyuan Allen-Zhu, Yuanzhi Li, and Zhao Song.
\newblock A convergence theory for deep learning via over-parameterization.
\newblock In Kamalika Chaudhuri and Ruslan Salakhutdinov, editors, {\em
  Proceedings of the 36th International Conference on Machine Learning},
  volume~97 of {\em Proceedings of Machine Learning Research}, pages 242--252.
  PMLR, 09--15 Jun 2019.

\bibitem{frankle2018the}
Jonathan Frankle and Michael Carbin.
\newblock The lottery ticket hypothesis: Finding sparse, trainable neural
  networks.
\newblock In {\em International Conference on Learning Representations}, 2019.

\bibitem{goodfellow2016deep}
Ian Goodfellow, Yoshua Bengio, and Aaron Courville.
\newblock {\em Deep learning}.
\newblock MIT press, 2016.

\bibitem{He2015}
Kaiming He, Xiangyu Zhang, Shaoqing Ren, and Jian Sun.
\newblock Deep residual learning for image recognition.
\newblock In {\em Proceedings of the IEEE conference on computer vision and
  pattern recognition}, pages 770--778, 2016.

\bibitem{hoefler2021sparsity}
Torsten Hoefler, Dan Alistarh, Tal Ben-Nun, Nikoli Dryden, and Alexandra Peste.
\newblock Sparsity in deep learning: Pruning and growth for efficient inference
  and training in neural networks.
\newblock {\em J. Mach. Learn. Res.}, 22(241):1--124, 2021.

\bibitem{Huang2017dcc}
Gao Huang, Zhuang Liu, Laurens Van Der~Maaten, and Kilian~Q. Weinberger.
\newblock Densely connected convolutional networks.
\newblock In {\em 2017 IEEE Conference on Computer Vision and Pattern
  Recognition (CVPR)}, pages 2261--2269, 2017.

\bibitem{kingma_adam_2014}
Diederik~P Kingma and Jimmy Ba.
\newblock {Adam: A method for stochastic optimization}.
\newblock {\em arXiv preprint arXiv:1412.6980}, 2014.

\bibitem{li2017vll}
Hao Li, Zheng Xu, Gavin Taylor, Christoph Studer, and Tom Goldstein.
\newblock Visualizing the loss landscape of neural nets, 2017.

\bibitem{lundby2022sdn}
Erlend Torje~Berg Lundby, Adil Rasheed, Jan~Tommy Gravdahl, and Ivar~Johan
  Halvorsen.
\newblock Sparse deep neural networks for modeling aluminum electrolysis
  dynamics.
\newblock {\em Applied Soft Computing}, 134:109989, February 2023.

\bibitem{robinson2022anc}
Haakon Robinson, Erlend Lundby, Adil Rasheed, and Jan~Tommy Gravdahl.
\newblock A novel corrective-source term approach to modeling unknown physics
  in aluminum extraction process.
\newblock {\em arXiv}, 2022.

\bibitem{WINTER2018802}
Maximilian Winter and Christian Breitsamter.
\newblock Nonlinear identification via connected neural networks for unsteady
  aerodynamic analysis.
\newblock {\em Aerospace Science and Technology}, 77:802--818, 2018.

\bibitem{Zhou2022sbd}
Hongpeng Zhou, Chahine Ibrahim, Wei~Xing Zheng, and Wei Pan.
\newblock Sparse {{Bayesian}} deep learning for dynamic system identification.
\newblock {\em Automatica}, 144:110489, October 2022.

\end{thebibliography}







\end{document}